\documentclass{article}
\usepackage[centertags]{amsmath}
\usepackage{amsfonts}
\usepackage{amssymb}
\usepackage{epsfig}
\usepackage[all]{xy}

 \newcommand{\Dscr}{\mathcal{D}}

\newcommand{\Hscr}{\mathcal{H}}

\newcommand{\Lscr}{\mathcal{L}}

\newcommand{\rmd}{{\mathrm{d}}}
\newcommand{\rmi}{{\mathrm{i}}}
\newcommand{\rme}{\mathrm{e}}

\newcommand{\Real}{\mathbb{R}}
\newcommand{\Complex}{\mathbb{C}}

\newcommand{\abs}[1]{\left\vert#1\right\vert}

\newcommand{\esssup}[1]{\operatorname{ess\;sup}}

\newcommand{\Tr}{\operatorname{Tr}}
\newcommand{\E}{\operatorname{\mathbb{E}}}

\begin{document}


\title{QUANTUM TRAJECTORIES, FEEDBACK AND SQUEEZING}

\author{A. Barchielli, M. Gregoratti, M. Licciardo
\\
Dipartimento di Matematica, Politecnico di Milano, \\
Piazza Leonardo da Vinci 32, I-20133, Milano, Italy}

\maketitle

\begin{abstract}
Quantum trajectory theory is the best mathematical set up to model continual observations of a
quantum system and feedback based on the observed output. Inside this framework, we study how
to enhance the squeezing of the fluorescence light emitted by a two-level atom, stimulated by
a coherent monochromatic laser. In the presence of a Wiseman-Milburn feedback scheme, based on
the homodyne detection of a fraction of the emitted light, we analyze the squeezing dependence
on the various control parameters.
\end{abstract}


\section{Introduction}

Photo-detection theory in continuous time has been widely developed
\cite{Dav76,Bar90QO,BarB91,BarPag96b,Bar05} and applied, in particular, to the fluorescence
light emitted by a two-level atom stimulated by a coherent monochromatic laser
\cite{WisMil93,Bar05}. As well as various feedback schemes on the atom evolution, based on the
outcoming photocurrent, have been proposed \cite{WisM93,Wis94,WMW02}. However the introduction
and the analysis of feedback have always been focused on the control of the atom dynamics. The
typical aim was to drive the atom to a preassigned asymptotic state or to a preassigned
asymptotic unitary dynamics \cite{WalM94,W98,WW00,WWM00,WMW02}.

Here, on the contrary, we are interested in the emitted light and in employing control and
feedback processes to enhance its squeezing properties. These can be checked by homodyne
detection and spectral analysis of the output current. For these reasons we consider the
mathematical description of photo-detection based on classical stochastic differential
equations (quantum trajectories), as it is suitable both to consistently compute the homodyne
spectrum of fluorescence light, and to introduce feedback and control in the mathematical
formulation. We study how the squeezing depends on the various control parameters and how
feedback mechanisms can be successfully introduced. We consider only Markovian feedback
schemes \`a la Wiseman-Milburn \cite{WisM93,Wis94}, as they leave the homodyne spectrum
explicitly computable.

\section{Detection and feedback scheme}

Consider a two-level atom with Hilbert space $\Hscr=\Complex^2$ and lowering and rising
operators $\sigma_-$ and $\sigma_+$. Let the Pauli matrices be $\sigma_x=\sigma_-+\sigma_+$,
$\sigma_y=\rmi(\sigma_--\sigma_+)$, $\sigma_z=\sigma_+\sigma_- - \sigma_-\sigma_+$ and let the
vector of operators $(\sigma_x,\sigma_y,\sigma_z)$ be denoted by $\vec\sigma$. Let also the
eigenprojectors of $\sigma_z$ be denoted by $P_+=\sigma_+\sigma_-$ and $P_-=\sigma_-\sigma_+$
and, for every angle $\phi$, let us introduce the unitary selfadjoint operator
\begin{equation*}
\sigma_\phi = \rme^{\rmi\phi}\,\sigma_- + \rme^{-\rmi\phi}\,\sigma_+ = \cos\phi\,\sigma_x+\sin\phi\,\sigma_y.
\end{equation*}
A state $\rho$ of the atom is represented by a point $\vec{x}$ in the Bloch sphere,
\begin{equation*}
\rho=\frac{1}{2}\left(1+\vec{x}\cdot\vec\sigma\right),\qquad\vec{x}\in\Real^3,\quad\abs{\vec{x}}\leq1.
\end{equation*}

We admit an open Markovian evolution for the atom, subjected to ``dephasing'' effects and to interactions both
with a thermal bath and, via absorption and emission of photons, with the electromagnetic field.
In particular we suppose that the atom is stimulated by a coherent monochromatic
laser and that the emitted light is partially lost in the so called \textit{forward channel} and partially
gathered in two so called \textit{side channels} for homodyne detection.

Let the free Hamiltonian of the atom be $\omega_0\sigma_z/2$, $\omega_0>0$. Let the natural
line-width of the atom be $\gamma$, let the intensities of the dephasing and thermal effects
be given by the adimensional parameters $k_\rmd\geq 0$ and $\overline n\geq 0$, let the
stimulating laser have frequency $\omega>0$ and Rabi frequency $\Omega\geq0$. Let
$\Delta\omega =\omega_0-\omega$ denote the detuning.

Let the fractions of light emitted in the forward and in the two side channels be $|\alpha_0|^2$,
$|\alpha_1|^2$, $|\alpha_2|^2$ respectively ($|\alpha_0|^2+|\alpha_1|^2+|\alpha_2|^2=1$, $|\alpha_0|^2>0$,
$|\alpha_1|^2>0$, $|\alpha_2|^2\geq0$), and, for the side channels, let the initial phase of the local
oscillator in the corresponding detector be $\vartheta_k=\arg\alpha_k$, $k=1,2$. Changing $\vartheta_k$ means
to change the measuring apparatus. Let the two homodyne photocurrents be $I_1$ and $I_2$.

We introduce a feedback control scheme \`a la Wiseman-Milburn based on the photocurrent $I_1$ revealed in
the side channel 1. Assuming instantaneous feedback, we modify the amplitude of the laser driving the atom by
adding a term $g\,\rme^{-\rmi\omega t}\,I_1(t)/\sqrt\gamma$ proportional to $I_1$, with the same frequency
$\omega$ and with initial phase possibly different from that of the original laser. Let this
difference be $\varphi$.
\[ \xymatrix{
& & & &
\\
*+[F]{\text{\begin{scriptsize}homodyne det.\end{scriptsize}}} \ar[u]^>>>{
I_2(t)}& &
*+[F-:<3pt>]{\text{\begin{scriptsize}atom\end{scriptsize}}}
\ar@{~>}[u]^>>>{\text{forward}}_>>>{\text{channel}}\ar@{~>}[rr]^{\text{side}}_{\text{channel 1}}
\ar@{~>}[ll]_{\text{side}}^{\text{channel 2}} & & *+[F]{\text{\begin{scriptsize}homodyne det.\end{scriptsize}}}
\ar@/^1pc/[dll]^<<<<<<<{
I_1(t)}
\\
&& *+[F]{\text{\begin{scriptsize}electro\-modulator\end{scriptsize}}}
\ar@{~>}[u]& & 
\\
&& \ar@{~} [u]_<<<{\text{laser}} & &}\]

Then the atom has a Markovian evolution, whether we condition its state on continuous
monitoring of the photocurrents, or we do not. Let us call \textit{a priori} state $\eta_t$
the unconditioned one and let us call \textit{a posteriori} state $\rho_t$ the conditioned
one. Of course $\eta_t$ is the mean of $\rho_t$. Let us write the evolution equations in the
rotating frame, where they result to be time-homogeneous. Let us introduce first the
parameters
\begin{equation*}
c=|g|\,|\alpha_0|/\sqrt\gamma\geq0, \qquad
\Delta\omega_c=\Delta\omega+c\,\gamma\,|\alpha_1|\cos(\vartheta_1-\varphi)\in\Real.
\end{equation*}
The a priori state $\eta_t$ is governed by the Master equation
\begin{equation*}
\rmd\eta_t= \Lscr\eta_t\,\rmd t,
\end{equation*}
\begin{multline*}
\Lscr\rho =
-\rmi\left[\frac{\Delta\omega_c}{2}\,\sigma_z
+ \frac{\Omega}{2}\,\sigma_x\;,\;\rho\right]
+ \gamma k_\rmd \left(\sigma_z\,\rho\,\sigma_z-\rho\right)
+\gamma\overline{n}\left(\sigma_+\,\rho\,\sigma_--\frac{1}{2}\left\{P_-\;,\;\rho\right\}\right)
\\
{}+\gamma(\overline{n}+1-|\alpha_1|^2)\left(\sigma_-\,\rho\,\sigma_+-\frac{1}{2}
\left\{P_+\;,\;\rho\right\}\right)
\\
{}+\gamma(\alpha_1\,\sigma_--\rmi
c\,\sigma_\varphi)\,\rho\,(\overline{\alpha}_1\,\sigma_++\rmi c\,\sigma_\varphi)
-\frac{\gamma}{2}\left\{\Big(|\alpha_1|^2-2c|\alpha_1|\sin(\vartheta_1-\varphi)\Big)P_++c^2\;,\;\rho\right\}.
\end{multline*}
The a posteriori state $\rho_t$ is governed by the non-linear stochastic Master equation
\begin{equation}\label{prior}
\rmd\rho_t= \Lscr\rho_t\,\rmd t + \sqrt\gamma \Dscr[\alpha_1\,\sigma_--\rmi c\,\sigma_\varphi]\rho_t\,\rmd
W_1(t) + \sqrt\gamma \Dscr[\alpha_2\,\sigma_-]\rho_t\,\rmd W_2(t),
\end{equation}
where, for every matrix $a$, the superoperator $\Dscr[a]$ is
\begin{equation*}
\Dscr[a]\rho=a\,\rho + \rho\,a^* - \rho \Tr\left[(a+a^*)\rho\right],
\end{equation*}
and where $W_1$ and $W_2$ are two independent Wiener processes. The two homodyne photocurrents
are given by the generalized stochastic processes
\begin{equation}\label{currents}
I_k(t)=\sqrt\gamma|\alpha_k| \Tr\left[\sigma_{\vartheta_k}\,\rho_t\right] + \dot{W}_k(t).
\end{equation}
Note that each signal term $\sqrt\gamma|\alpha_k| \Tr\left[\sigma_{\vartheta_k}\,\rho_t\right]$
depends on the dynamics of the a posteriori state $\rho_t$ and that, typically, it is
correlated to both the noise terms $\dot{W}_1(t)$ and $\dot{W}_2(t)$.

Let us remark that, even if the feedback is based on the singular stochastic process $I_1$,
the mathematical formulation of the model is not affected by this singularity, as we do not
observe directly the light in the forward channel.

We suppose that $|\alpha_0|$ is assigned by experimental constraints and that the control
parameters are $\Omega$, $\Delta\omega$, $|\alpha_1|$, $|\alpha_2|$, $\vartheta_1$,
$\vartheta_2$, $c$ and $\varphi$. Of course, if $c=0$, then there is no feedback action on the
atom, so that its a priori dynamics is independent of the measurement process, that is of the
fractions $|\alpha_1|^2$, $|\alpha_2|^2$ and of the phases $\vartheta_1$, $\vartheta_2$, $\varphi$.
On the contrary, if $c>0$, then the
a priori dynamics is modified by the feedback loop and it depends also on $|\alpha_1|$,
$\vartheta_1$, $c$ and $\varphi$.

\section{Homodyne incoherent spectrum and squeezing}

We are interested in the light emitted by the atom and in particular in the squeezing
properties of the light in the side channels 1 and 2. With the help of the incoherent spectrum
of the homodyne photocurrents we can analyze the squeezing properties of the light detected in
the two side channels, and thus we can investigate the effect of the control parameters.

When $|\alpha_2|^2=0$, the fluorescence light which is not lost in the forward channel is
gathered in a unique side channel, so that the squeezing is analyzed just for that light which
is also detected for the feedback loop. This means that the eventually squeezed light would
not be available for other purposes. Thus in this case a unique homodyne detector is employed
and $|\alpha_1|^2$ is its efficiency. The meaning and the possible usefulness of the squeezing
of the light involved in the feedback loop is discussed by Wiseman \cite{W98}. When
$|\alpha_2|^2>0$, the fluorescence light which is not lost in the forward channel is split in
the two side channels. The homodyne detection of the light in channel 1 allows both the
analysis of its squeezing and the feedback control. The light in channel 2 is detected for
squeezing analysis, as well as it could be employed for different uses. Let us remark that,
even if we were interested in squeezing only for channel 2, the choice of a feedback scheme
based on homodyne detection in channel 1 would be still essential in order to get a
time-homogeneous atomic evolution in the rotating frame.

The structure of $\Lscr$ guarantees that, for every initial preparation of the atom, the a priori state
$\eta_t$ asymptotically reaches the stationary state
\begin{equation}\label{equilibrium}
\rho_\textrm{eq}=\frac{1}{2}\left(1+\vec{x}_\textrm{eq}\cdot\vec\sigma\right), \qquad
\vec{x}_\textrm{eq} =
-\gamma\Big(1-2c|\alpha_1|\sin(\vartheta_1-\varphi)\Big)\,A^{-1}\,
\begin{pmatrix}0\\0\\1\end{pmatrix},
\end{equation}
where $A$ is the $3\times3$ matrix giving $\Lscr$ in the Bloch sphere language,
\begin{gather*}
A=\begin{pmatrix}a_{11}&a_{12}&0\\
a_{21}&a_{22}&\Omega\\
0&-\Omega&a_{33}\end{pmatrix},\\
a_{11}=\gamma\Big(\frac{1}{2}+\overline{n}+2k_\rmd + 2c|\alpha_1|\cos\vartheta_1\sin\varphi+2c^2\sin^2\varphi\Big),\\
a_{12}=\Delta\omega_c - \gamma\Big(c|\alpha_1|\cos(\vartheta_1+\varphi)-2c^2\sin2\varphi\Big),\\
a_{21}=-\Delta\omega_c - \gamma\Big(c|\alpha_1|\cos(\vartheta_1+\varphi)-2c^2\sin2\varphi\Big),\\
a_{22}=\gamma\Big(\frac{1}{2}+\overline{n}+2k_\rmd - 2c|\alpha_1|\sin\vartheta_1\cos\varphi+2c^2\cos^2\varphi\Big),\\
a_{33}=\gamma\Big(1+2\overline{n} - 2c|\alpha_1|\sin(\vartheta_1-\varphi)+2c^2\Big).
\end{gather*}
Thus we can introduce the homodyne incoherent spectrum of the light revealed in each side channel $k$ as the
limit of the normalized variance of the Fourier transform of the photocurrent $I_k$
\begin{equation*}
S_k(\mu)=\lim_{T\to+\infty}\frac{1}{T}\left\{\E\left[\abs{\int_0^T\rme^{\rmi\mu s}\,I_k(s)\,\rmd s}^2\right] -
\abs{\E\left[\int_0^T\rme^{\rmi \mu s}\,I_k(s)\,\rmd s\right]}^2\right\}.
\end{equation*}
It is a positive even function of its real argument $\mu$ which can be computed from equations
\eqref{prior} and \eqref{currents} by Ito calculus and by the full theory of Quantum Continual
Measurement, which can provide the first and second moments of $I_1$ \cite{Bar90QO,Bar05}.
Thus, for every initial state of the atom, we can obtain
\begin{equation}\label{spectrum}
S_k(\mu)=1+2\gamma|\alpha_k|^2\,\left(\frac{A}{A^2+\mu^2}\,\vec{t}_k\right)\cdot\vec{s},
\end{equation}
where $\vec{t}_k$ and $\vec{s}$ are the vectors in $\Real^3$ defined as
\begin{gather*}
\vec{t}_1=\Tr\Big[\big(\rme^{\rmi\vartheta_1}\,\sigma_-\,\rho_\textrm{eq} +
\rme^{-\rmi\vartheta_1}\,\rho_\textrm{eq}\,\sigma_+
- \Tr[\sigma_{\vartheta_1}\,\rho_\textrm{eq}]\,\rho_\textrm{eq} +
\rmi\frac{c}{|\alpha_1|}\,[\rho_\textrm{eq},\sigma_\varphi]\big)\,\vec\sigma\Big],\\
\vec{t}_2=\Tr\Big[\big(\rme^{\rmi\vartheta_2}\sigma_-\,\rho_\textrm{eq} +
\rho_\textrm{eq}\,\rme^{-\rmi\vartheta_2}\sigma_+
- \Tr[\sigma_{\vartheta_2}\,\rho_\textrm{eq}]\,\rho_\textrm{eq}\big)\,\vec\sigma\Big],\\
\vec{s}=\begin{pmatrix}\cos\vartheta_k\\\sin\vartheta_k\\0\end{pmatrix}.
\end{gather*}
More explicitly, by using the Bloch components of the equilibrium state \eqref{equilibrium}, we get
\[
\vec{t}_1=\begin{pmatrix}\left(1+z_\textrm{eq}-x_\textrm{eq}^{\;2}\right)
\cos\vartheta_1-x_\textrm{eq} y_\textrm{eq} \sin\vartheta_1\\
\left(1+z_\textrm{eq}-y_\textrm{eq}^{\;2}\right) \sin\vartheta_1-x_\textrm{eq} y_\textrm{eq}
\cos\vartheta_1 \\
-\left(1+z_\textrm{eq}\right) \left(x_\textrm{eq}\cos\vartheta_1+ y_\textrm{eq}
\sin\vartheta_1\right)\end{pmatrix} + \frac {2c}{|\alpha_1|} \begin{pmatrix} z_\textrm{eq}
\sin \varphi \\ - z_\textrm{eq} \cos \varphi \\  -x_\textrm{eq} \sin \varphi +y_\textrm{eq}
\cos \varphi \end{pmatrix},
\]
\[
\vec{t}_2=\begin{pmatrix}\left(1+z_\textrm{eq}-x_\textrm{eq}^{\;2}\right)
\cos\vartheta_2-x_\textrm{eq} y_\textrm{eq} \sin\vartheta_2\\
\left(1+z_\textrm{eq}-y_\textrm{eq}^{\;2}\right) \sin\vartheta_2-x_\textrm{eq} y_\textrm{eq}
\cos\vartheta_2 \\
-\left(1+z_\textrm{eq}\right) \left(x_\textrm{eq}\cos\vartheta_2+ y_\textrm{eq}
\sin\vartheta_2\right)\end{pmatrix}.
\]
Each spectrum $S_k$ depends on $k_\rmd$, $\overline{n}$, $\Omega$, $\Delta\omega$,
$|\alpha_k|$, $\vartheta_k$, $c$ and $\varphi$. Moreover, $S_2$ depends also on $|\alpha_1|$
and $\vartheta_1$.

If $S_k(\mu)<1$ for some $\mu$ and $\vartheta_k$, then the homodyne detection identified by
$\vartheta_k$ reveals a squeezed mode around $\mu$ of the light in channel $k$.

Independently of the presence of the feedback loop, every time a parameter $|\alpha_k|$
vanishes, the corresponding photocurrent $I_k$ reduces to a pure white noise (shot noise due
to the local oscillator) with spectrum $S_k=1$ for every choice of the other parameters.

Analyze first $c=0$, the situation without
feedback. In this case each dependence on $\varphi$ disappears and $S_2$ becomes independent
of $|\alpha_1|$ and $\vartheta_1$, so that there is no difference between $S_1$ and $S_2$.
Moreover, the dependence of each spectrum $S_k$ on the corresponding $|\alpha_k|$ reduces to the explicit
multiplication coefficient in \eqref{spectrum}. Therefore, when the control parameters $\Omega$ and
$\Delta\omega$ give squeezed light in channel $k$, the lowering of $S_k$ under the shot noise level is anyhow
directly proportional to the fraction of emitted light gathered in that channel.

For $\Omega=0$ and $\overline n=0$ there is no
fluorescence light in the long run, so that each photocurrent $I_k$ again
reduces to a pure white noise with spectrum $S_k=1$.

For $\Omega=0$ and $\overline n>0$ there is no dependence on $\vartheta_k$ and $S_k>1$.
In this case there is only thermal light with carrier frequency $\omega_0$, while
the local oscillator is at frequency $\omega$. The result are two temperature
dependent Lorentzian peaks at $\mu=\pm \Delta \omega$. The white noise contribution
is always present.

When $\Omega>0$, $S_k$ becomes $\vartheta_k$-dependent and it can go below the shot noise
level. This fact means that some negative correlation between signal and noise has been
developed. Some examples are plotted for both channels, always for $\gamma=1$, $k_\rmd =0$,
$\overline n=0$ and $|\alpha_1|^2=|\alpha_2|^2=0.45$. Figures 1 and 2 show $S_1$ and $S_2$
respectively for $\Delta\omega=0$ (line 1) and for $\Delta\omega=-2$ (line 3), every time for
values of $\Omega$ and $\vartheta_k$ chosen in order to have a region with a pronounced
squeezing ($\Omega=0.2976$, $\vartheta_k=-\pi/2$ for line 1; $\Omega=2.0526$,
$\vartheta_k=0.1449$ for line 3).

One could also compare the homodyne spectrum with and without $\overline n$ and $k_\rmd$, thus
verifying that the squeezing is very sensitive to any small perturbation.

\begin{figure}[b]
\begin{center}
\parbox{5.5cm}{%
\psfig{file=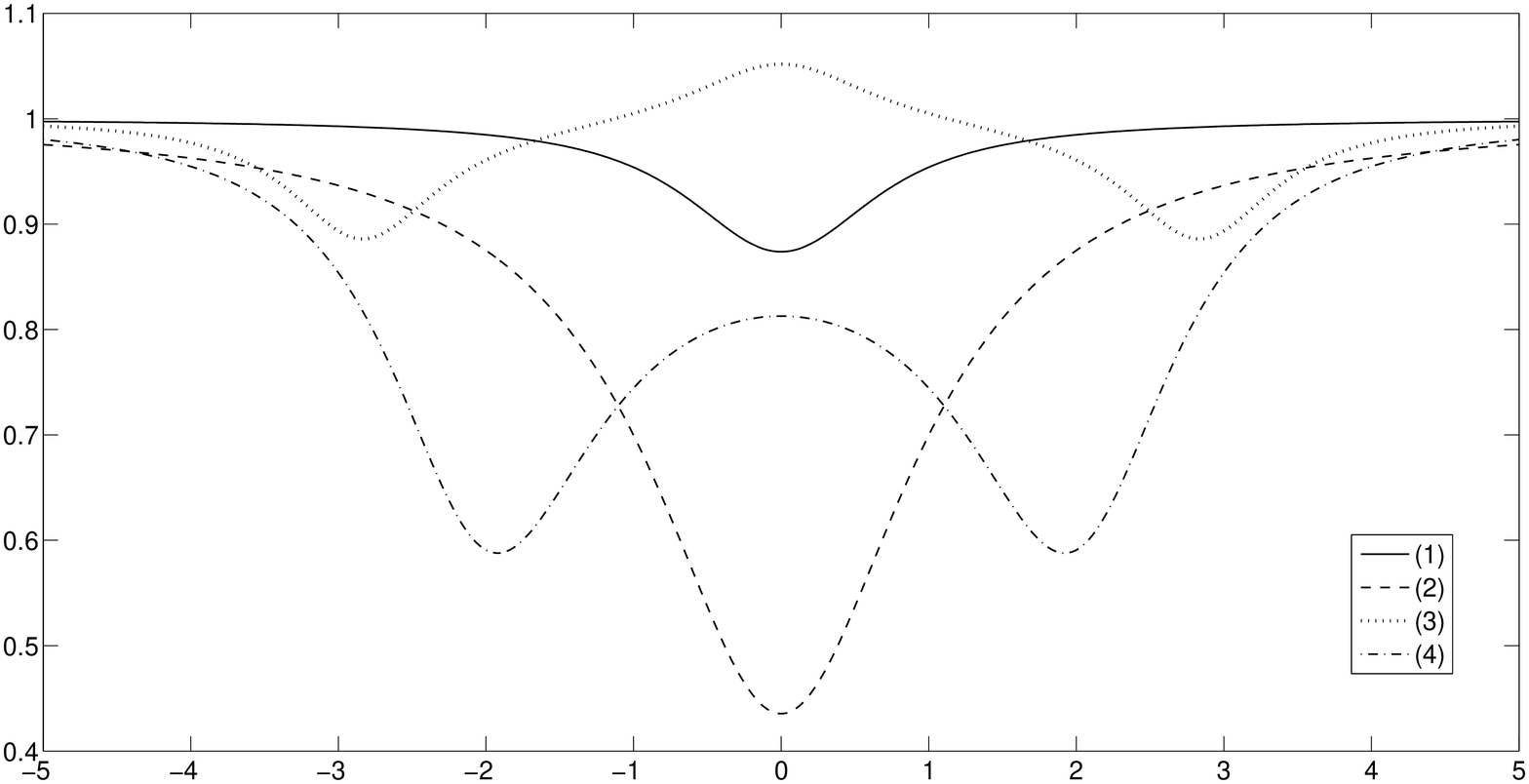,height=4cm,width=5.5cm} \caption{Channel 1} \label{fig1}} \qquad
\begin{minipage}{5.5cm}
\psfig{file=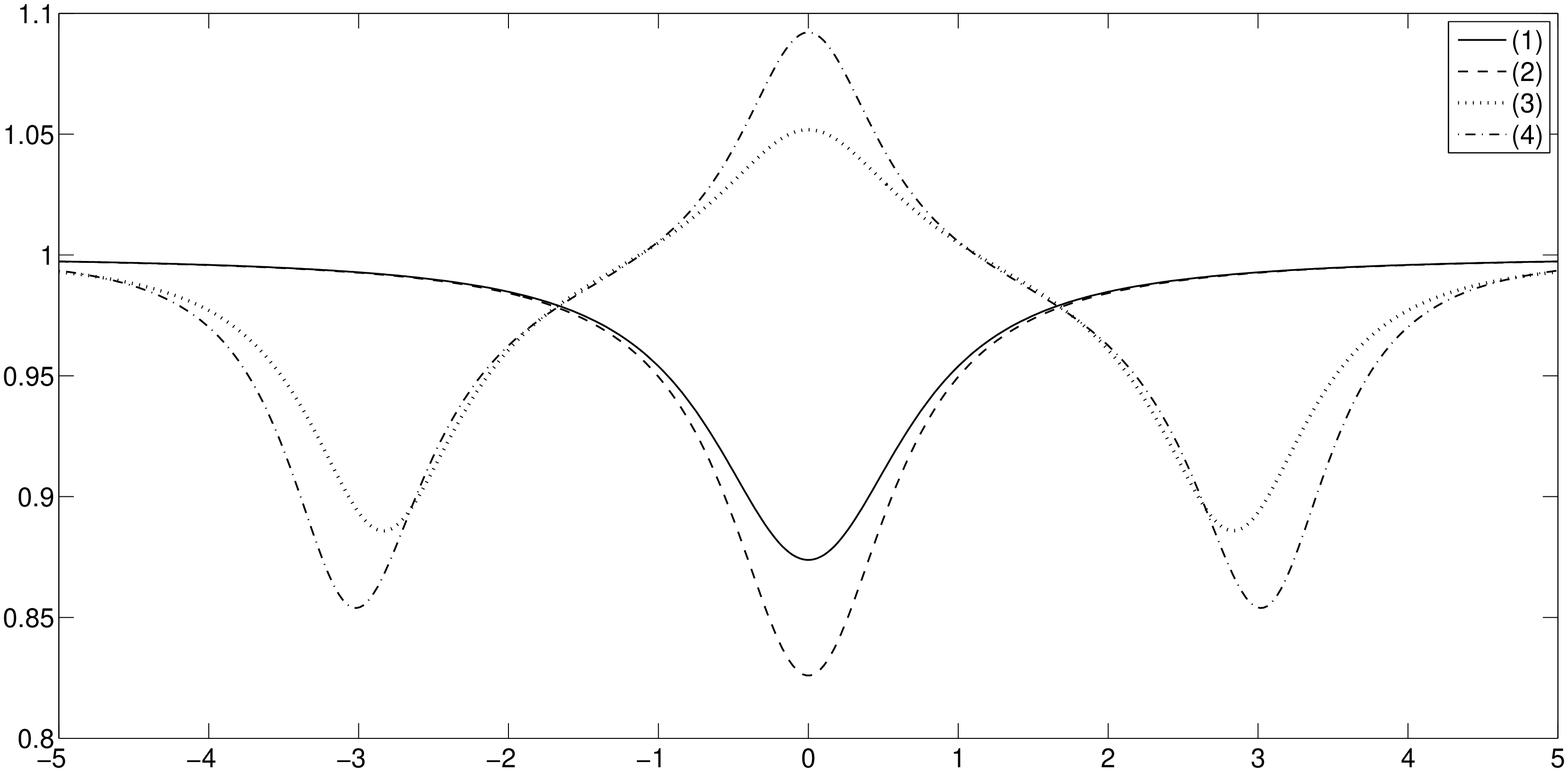,height=4cm,width=5.5cm} \caption{Channel 2} \label{fig2}
\end{minipage}
\end{center}
\end{figure}

Allow now $c\geq0$. The optimal squeezing in channel 1 is always found for $\Omega^2=0$ and
the feedback loop is very helpful, giving good visible minima of $S_1$ also when $|\alpha_1|^2$ is not
close to 1. For example, in the case $|\alpha_1|^2=0.45$, Fig.\ 1 shows $S_1$ for
$\Delta\omega=0$ (line 2) and for $\Delta\omega=-2$ (line 4), every time for values of
$\Omega$, $\vartheta_1$, $c$ and $\varphi$ chosen in order to enhance the squeezing
($\Omega=0$, $c=0.2936$, $\varphi-\vartheta_1=\pi/2$ for line 2; $\Omega=0$, $c=0.3762$,
$\vartheta_1=0.0482$, $\varphi=1.9941$ for line 4). Again $\gamma=1$, $k_\rmd =0$, $\overline
n=0$. The utility of the feedback scheme can be appreciated by comparing lines 1 and 3 with
lines 2 and 4 respectively.

If we are interested in the light emitted in channel 2 and if $|\alpha_1|^2$ and
$|\alpha_2|^2$ are assigned by some constraints, then the squeezing in channel 2 can be
enhanced by a feedback scheme based on the photocurrent coming from channel 1, but the
feedback performance is not as good as it can be for the squeezing in channel 1 itself. Fig.\ 2
shows $S_2$ for $\Delta\omega=0$ (line 2) and for $\Delta\omega=-2$ (line 4), every time for
values of $\Omega$, $\vartheta_1$, $c$, $\varphi$ and $\vartheta_2$, chosen in order to
enhance the squeezing ($\Omega=0.2698$, $\vartheta_1=\pi/2$, $c=0.0896$, $\varphi=0$,
$\vartheta_2=\pi/2$ for line 2; $\Omega=2.329$, $\vartheta_1=0.2896$, $c=0.1346$,
$\varphi=-1.2902$, $\vartheta_2=0.0728$ for line 4).

Anyway, if the only constraint is $|\alpha_1|^2 + |\alpha_2|^2 = 1-|\alpha_0|^2$ and we are
free in the choice of $|\alpha_1|^2$ and $|\alpha_2|^2$, then the best observable squeezing in
channel 2 is obtained in the limit case $|\alpha_1|^2=0$, $c=0$. That is, when the whole
non-lost light is gathered just in channel 2 and the white noise $I_1$ revealed in channel 1
is ignored.

Let us remark that, when we use control parameters enhancing the squeezing for channel $k$,
every time $\Delta\omega=0$ we find the spectrum $S_k$ with an absolute minimum in $\mu=0$,
while whenever $\Delta\omega\neq0$ we find the spectrum $S_k$ with two absolute minima,
symmetric with respect to $\mu=0$, which turns out to be a local maximum.

Finally let us remark that the idea of the papers \cite{WW00,WWM00,WMW02} was to choose the
control parameters in such a way that, in the rotating frame, the atom is frozen in a
preassigned pure state $h_0\in\Hscr$, i.e.\ in such a way that, in the rotating frame, both
the a priori state $\eta_t$ and the a posteriori state $\rho_t$ asymptotically reach
$\rho_\textrm{eq}=|h_0\rangle\langle h_0|$. This is possible in an exact way only in a very
ideal case, which in our notations corresponds to $\Delta\omega=0$, $\varphi=0$,
$\vartheta_1=\pm\pi/2$, $|\alpha_1|=1$, $\alpha_0=\alpha_2=0$, $k_\rmd=0$, $\overline n=0$,
which implies in particular $a_{12}=a_{21}=0$ and $x_\textrm{eq}=0$. Then, the a posteriori
state $\rho_t$ is driven to a pure given state if $\Omega$ and $c$ are such that
$y_\textrm{eq}^{\;2}+z_\textrm{eq}^{\;2}=1$ and $2c \sin \vartheta_1=1+z_\textrm{eq}$. But
this implies $\vec{t}_1=0$ and the two incoherent spectra reduce to pure shot noise. This is
reasonable: if the atom if frozen there is not incoherent scattering of light. One can check
that actually only the coherent scattering survives, giving a $\delta$-contribution in $\mu=0$
to the complete spectrum. If the freezing of the atom is only approximate, or if one tries to
maximize the atomic squeezing (which is another way of stopping the atomic motion), one can
check that all the spectra tend to become flatter and the squeezing tends to disappear.

\vspace*{-6pt}   

\end{document}